\documentstyle[aps,twocolumn,epsf]{revtex}

\begin{document}

\draft

\title{
Ideal Quantum Communication over Noisy Channels: a Quantum Optical
Implementation
}

\author{S.J. van Enk$^{(1,2)}$, J.I. Cirac$^{(1,2)}$, 
and P. Zoller$^{(2)}$}

\address{
(1) Departamento de Fisica, Universidad de
Castilla--La Mancha, 13071 Ciudad Real, Spain.
}

\address{
(2) Institut f\"ur Theoretische Physik, Universit\"at Innsbruck,
Technikerstrasse 25, A--6020 Innsbruck, Austria.
}

\date{\today}

\maketitle

\begin{abstract}
We consider transmission of a quantum state between two distant atoms
via photons. Based on a quantum-optical realistic model, we define a
noisy quantum channel which includes systematic errors as well as
errors due to coupling to the environment. We present a protocol that
allows one to accomplish ideal transmission by repeating the transfer
operation as many times as needed.
\end{abstract}


\narrowtext

Quantum communication \cite{QCrypto,Schumacher,Purification,densec}
is the transmission and exchange of quantum
information between distant ``nodes'' of a quantum network. 
The nodes of a
quantum network are typically two-level atoms which store the quantum
information represented by entangled states of quantum bits
(qubits). Operations in such a quantum network are unitary
transformations on qubits. These can either be local operations,
i.e.\ within a node, or non-local operations involving qubits in
distant nodes, such as transmission of qubits or, in general,
distribution of entanglement over the network. In particular, ideal
quantum transmission is defined by
\begin{equation}
(c_0 |0\rangle_1 + c_1 |1\rangle)\otimes |0\rangle_2
\longrightarrow 
|0\rangle_1 \otimes (c_0 |0\rangle_2 + c_1 |1\rangle_2).
\end{equation}
where an  unknown
superposition of internal states $|0\rangle$ and $|1\rangle$ in atom 1
in node 1 is transfered to atom 2 in node 2.

Physical implementations of transmission protocols in a quantum
network based on cavity QED (CQED) have recently been proposed
\cite{QN}. They involve properly designed laser pulses which excite an
atom inside an optical cavity at the sending node, so that the state
is mapped into a photon wavepacket. This wavepacket propagates along a
transmission line connecting the cavities, enters the cavity at the
receiving node, and is absorbed by an atom [See Fig.\ 1(a)]. In other
words, ``permanent qubits'' stored in atoms generate and annihilate
``transient qubits'' represented by photons which play the role of a
data bus for quantum information. In a perfect implementation, this
scheme allows for ideal transmission.  In practice, there will be
errors; in particular, the channel through which the photon travels
will be noisy, i.e.\ in the transmission between distant nodes there
will be, for example, photon absorption. In this Letter we will show
that this physical setting of quantum transmission of qubits via
photon exchange gives rise to novel error correction schemes to be
developed, which permit one to retry sending the qubit until perfect
transmission is achieved, thus correcting for transmission errors in
the noisy quantum optical channel to {\em all orders}. Typical error
correction schemes developed in the context of quantum computing
\cite{errorcor} perform redundant encoding of a logical qubit in
several physical qubits, i.e.\ add an increasing number of atoms
(``permanent qubits''), to correct errors to higher order
\cite{qcerror}.  In contrast, the physical basis of our scheme, laser
manipulation of atoms in CQED \cite{Kimble}, makes it comparatively
simple to create highly entangled states of atoms with many photons
(``transient qubits''). Furthermore, these correction schemes for
transmission errors require only a moderate overhead, with
(entanglement of) only two atoms on the sending side, and two atoms on
the receiving node. Given that small model systems of ``ion trap
quantum computers'' \cite{Iontrap} involving a few qubits will be
built in the near future, such a scheme opens a realistic perspective
of implementing perfect transmission in quantum networks. This will
have interesting applications, such as distributing and storing EPR
pairs (or $N$--atom entangled states) in distant nodes for secure
public key distribution \cite{QCrypto}, purification schemes for
quantum cryptography \cite{Purification}, and dense coding of quantum
information \cite{densec}.
\begin{figure}[htbp]
  \begin{center}
   \leavevmode
      \epsfxsize=8cm  \epsfbox{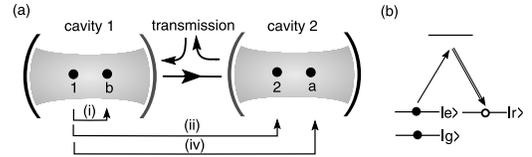}  
\caption{ (a) Schematic representation of the protocol for ideal
transmission of one qubit from atom 1 in cavity 1 to atom 2 in cavity
2. Atom $b$ is a backup atom and $a$ an auxiliary atom.  The steps
(i)--(v) in the protocol can be found in the text.  (b) Level
structure of atoms and couplings induced by laser and cavity fields.
}
  \end{center}
\end{figure}
We consider the atomic scheme outlined in Fig.~1(b). Three internal
long--lived (ground) states levels participate in the transmission.
The qubit is stored in $|g\rangle$ and $|e\rangle$, whereas
$|r\rangle$ acts as an auxiliary level. To achieve transmision from
atom $i$ to $j$, one first transfers $|e\rangle_i \rightarrow
|r\rangle_i$ via a Raman process, where a photon is emitted into a
high--Q cavity.  The generated photon leaks out of the cavity,
propagates along the transmission line, enters the optical cavity at
the second node, and induces the inverse transition $|R\rangle_j
\rightarrow |E\rangle_j$ (we will use capital letters to denote the
states of the atoms in the second node). In the ideal case, for time
$t\rightarrow \infty$ this corresponds to a mapping of the atomic
states
\begin{equation}
\label{Gijid}
{\cal T}_{i,j}:
\begin{array}{lll}
|g\rangle_i|R\rangle_j&\rightarrow& |g\rangle_i|R\rangle_j\\
|e\rangle_i|R\rangle_j&\rightarrow& |r\rangle_i|E\rangle_j,
\end{array}
\end{equation}
where the modes of the electromagnetic field are restored to the vacuum.

In reality, there will be errors due to coupling to the environment,
as well as systematic errors due to imperfections in adjusting the
experimental parameters. We consider the errors that occur during the
transmission, and assume that local operations are error-free.  The
most important errors during transmission will be due to: (1) photon
absorption, either in the mirrors or in the transmission line; (2) 
imperfectly designed laser pulses (including timing, detuning,
etc.); the photon wavepacket may be reflected from the second cavity
or induce incorrect transfer in the second atom; (3) uncontrolled
phase shifts and polarization changes in the transmission line; (4)
spontaneous emission during the Raman process. This last error can be
strongly suppressed by detuning the laser and the cavity mode from the
excited atomic states.  Nevertheless, we allow for spontaneous
emission to the states $|r\rangle_i$ or $|R\rangle_j$
\cite{Interf}. 

Decoherence and decay may be viewed as a result of a coupling between
the system (the two nodes) and the environment.  Under the assumption
of vanishing correlation time for the reservoir (Markov
approximation), the time evolution of the system can be described by a
pure state vector evolving according to a nonhermitian effective
Hamiltonian ($H_{\rm eff}$) interrupted by quantum jumps at random
times. This quantum jump picture of dissipative dynamics underlines
the recently developed quantum trajectories methods developed from
Monte Carlo integration of quantum optical master equations
\cite{review}.  More specifically, our present setting [Fig.~1(a)]
corresponds to a cascaded quantum system where there is a {\em
unidirectional} coupling from the first to the second node. The
general theory of cascaded quantum systems, in particular the quantum
trajectory formulation, was developed by Carmichael and Gardiner
\cite{cascade}. Systematic errors are included in this description as
part of the effective Hamiltonian evolution.  Within the present model
there are two possible evolutions during nonideal transmission, which
can be summarized as follows:

(i) With a probability $P$, no jump will occur.
The corresponding evolution will be given by 
\begin{equation}
\label{nojump}
\begin{array}{l}
|g\rangle_i |R\rangle_j \\
|e\rangle_i |R\rangle_j 
\end{array}
\stackrel{H_{\rm eff}}{\longrightarrow} 
\begin{array}{l}
\alpha|g\rangle_i |R\rangle_j \\
\beta |r\rangle_i |E\rangle_j + \gamma_1  |r\rangle_i |R\rangle_j
+\gamma_2  |e\rangle_i |R\rangle_j
\end{array}
\end{equation}
for $t\rightarrow \infty$,
where we do not write the state of the cavity modes explicitly since
it starts and ends up in the vacuum state $|00\rangle_c$.  The
appearance of population in levels $|r\rangle_i |R\rangle_j$ and
$|e\rangle_i |R\rangle_j$ may be due to wrongly designed laser
pulses. There can also be phase shifts and amplitude damping of the
coefficients $\alpha$ and $\beta$, for example, due to photon
absorption or spontaneous emission. In general, the complex
coefficients $\alpha,\beta$ and $\gamma_{1,2}$ will be functions of
(random) external parameters. We will, however, assume for a given
complete process (i)--(v) [see Fig.~1(a) and below], $\alpha$ and
$\beta$ are the same in the first (ii) and second (iv) transmission
\cite{lpulses}.

(ii) With a probability $1-P$ a quantum jump will occur corresponding
to either photon absorption or spontaneous emission from one of the
atoms. The complete process will consist of an evolution according to
the effective Hamiltonian, a quantum jump at a random time $\tau$, and
followed by the evolution given by $H_{\rm eff}$. For $t\rightarrow
\infty$, this can be summarized as
\begin{equation}
\label{jump}
\begin{array}{l}
|g\rangle_i |R\rangle_j \\
|e\rangle_i |R\rangle_j 
\end{array}
\stackrel{qj}{\longrightarrow} 
\begin{array}{l}
0 \\
|r\rangle_i |R\rangle_j
\end{array}
\end{equation}
where the cavity modes are again restored to the vacuum state.
Physically, Eq.~(\ref{jump}) can be understood as follows: If a photon
is absorbed while propagating from the first to the second cavity, or
in the cavity mirrors, this means that it was emitted by atom 1, which
ends up in state $|r\rangle_i$; atom 2 remains in $|R\rangle_j$, since
there is no photon to excite it. In a similar way, if the first atom
undergoes spontaneous emission to level $|r\rangle_i$ during the Raman
process, no photon will be transmitted via the channel, and again atom
2 will remain in state $|R\rangle_j$. The same reasoning applies to
spontaneous emission in atom 2 to level $|R\rangle_j$.  Note that
(\ref{jump}) is a special case of the state mapping (\ref{nojump})
with $\alpha=\beta=\gamma_2=0$.

We can summarize and formalize the above discussion in the following
definition of a noisy channel. Consider the state mapping defined in
(\ref{nojump}): 

\begin{itemize}
\item With probability $P\neq 0$, $\alpha,\beta,$ and $\gamma_{1,2}$
are random constants, but $\alpha,\beta$
are the same in two consecutive transmissions
[(ii) and (iv) in Fig.~1(a) and see below].

\item With probability $1-P$, $\alpha=\beta=\gamma_2=0$. 
\end{itemize}

Now we show how to perform ideal transmission over this noisy channel,
for arbitrarily small $P$. In the following, normalization factors are
left out.
We start out with the superposition 
$(c_0|g\rangle_1 + c_1 |e\rangle_1) |g\rangle_b|R\rangle_2|R\rangle_a$. 
The
scheme consists of five steps [Fig.~1(a)]:

{\em (i) Local redundant encoding:} 
Entangle atom 1 with the backup atom $b$ in node 1:
\[
\left[\!\begin{array}{l} |g\rangle_1 \\ |e\rangle_1 \end{array}\!\right]
\!\otimes\! |g\rangle_b|R\rangle_2|R\rangle_a
\rightarrow
\left[\!\begin{array}{l} |g\rangle_b \\ |e\rangle_b \end{array}\!\right]
\!\otimes \!|\Psi_i\rangle +
\left[\!\begin{array}{l} |e\rangle_b \\ |g\rangle_b \end{array}\!\right]
\!\otimes\! |\Phi_i\rangle.
\]
where
\begin{equation}
|\Psi_i\rangle = |e\rangle_1|R\rangle_2 |R\rangle_b,\quad
|\Phi_i\rangle = |g\rangle_1|R\rangle_2 |R\rangle_b.
\end{equation}
In the rest of the scheme atom $b$ will not participate in any process. 
Thus, we just have to give the evolution of the
states $|\Psi_i\rangle$ and $|\Phi_i\rangle$. 

{\em (ii) Transmission from atom 1 to 2:} We find
\begin{mathletters}
\begin{eqnarray}
\label{ii}
|\Psi_{ii}\rangle &=& \left(\beta |r\rangle_1 |E\rangle_2 + \gamma_1  
|r\rangle_1 |R\rangle_2
+\gamma_2  |e\rangle_1 |R\rangle_2\right)\, |R\rangle_a,\\
|\Phi_{ii}\rangle &=& \alpha|g\rangle_1 |R\rangle_2 |R\rangle_a.
\end{eqnarray}
\end{mathletters}
Then we measure if atom 1 is left in state $|e\rangle$. If yes, an
error has occurred, and the state is collapsed to
\[
\left[\!\begin{array}{l} |g\rangle_b \\ |e\rangle_b \end{array}\!\right]
\!\otimes\! |e\rangle_1|R\rangle_2|R\rangle_a.
\]
The backup atom is in the pure state $c_0|g\rangle_b+c_1|e\rangle_b$,
so that we can start again, after resetting the remaining atoms. If
atom 1 is not found in state $|e\rangle_1$, the corresponding
component in (\ref{ii}) is projected out.

{\em (iii) Symmetrization:} 
Perform a local operation on atom 1 that takes $|r\rangle_1$ to
$|g\rangle_1$, and $|g\rangle_1$ to $|e\rangle_1$, so that
\begin{mathletters}
\begin{eqnarray}
|\Psi_{iii}\rangle &=& \left(\beta |g\rangle_1 |E\rangle_2 + \gamma_1  
|g\rangle_1 |R\rangle_2 \right) \, |R\rangle_a,\\
|\Phi_{iii}\rangle &=& \alpha|e\rangle_1 |R\rangle_2 |R\rangle_a,
\end{eqnarray}
\end{mathletters}
By effectively interchanging $|g\rangle_1$ and $|e\rangle_1$, the 
unknown coefficients $\alpha$ and $\beta$ of the first transmission
are now ``symmetrized'': in the next step,
$|\Phi\rangle$ will acquire exactly those
phase and amplitude errors which 
$|\Psi\rangle$ acquired in step (ii) \cite{TP1}.

{\em (iv) Transmission from atom 1 to $a$:} We obtain
\begin{eqnarray*}
|\Psi_{iv}\rangle &=& \tilde\alpha\beta |g\rangle_1 |E\rangle_2 |R\rangle_a
+ \tilde\alpha \gamma_1 |g\rangle_1 |R\rangle_2 |R\rangle_a,\\
|\Phi_{iv}\rangle &=& \alpha \tilde\beta|r\rangle_1 |R\rangle_2 |E\rangle_a
+ \alpha (\tilde \gamma_1 |r\rangle_1 + \tilde \gamma_2 |e\rangle_1)
|R\rangle_2 |R\rangle_a,
\end{eqnarray*}
where the $\tilde\alpha$ etc. refer to the second transmission. Then
we measure if atom 1 is in $|e\rangle_1$. If yes, an error has
occurred and the state of atom $b$ can be recovered similar to step
(ii). If not, measure if atoms 2 and $a$ are in
$|R\rangle_2|R\rangle_a$. If yes, an error has occurred and we measure
the state of atom 1. Depending on the outcome, an appropriate one--bit
operation allows us to recover the state from the backup atom. If
atoms 2 and $a$ are not found in $|R\rangle_2|R\rangle_a$, then it
implies that no quantum jump has occurred, and therefore, according to
our assumption $\alpha=\tilde\alpha$ and $\tilde\beta=\beta$,
the unknown coefficients $\alpha$ and $\beta$ factorize, and
thus drop out. The states will be now
\begin{equation}
\label{tp}
\left[\!\begin{array}{l} |g\rangle_b \\ |e\rangle_b \end{array}\!\right]
\!\otimes \! |g\rangle_1 |E\rangle_2 |R\rangle_a+
\left[\!\begin{array}{l} |e\rangle_b \\ |g\rangle_b \end{array}\!\right]
\!\otimes\! |r\rangle_1 |R\rangle_2 |E\rangle_a.
\end{equation}

{\em (v) Teleportation:} We measure whether atom $b$ is in
$|g\rangle_b$ or in $|e\rangle_b$. Then we measure if atom 1 is in
$|g\rangle_1 \pm |e\rangle_1$.  Finally, measure if atom $a$ is in
$|E\rangle_1 \pm |R\rangle_1$.  Depending on the outcome of these
measurements, one can apply an appropriate single atom operation to
atom 2 to obtain the original superposition $c_0 |G\rangle_2 + c_1
|E\rangle_2$ with probability one. These measurements effectively
teleport the state from the first to the second node \cite{DVNZ}.

We now present numerical simulations of the full problem, to
illustrate this error scheme in the context of quantum Monte Carlo
wave function simulations for a cascaded quantum system. We take as
the effective Hamiltonian for our system (in the rotating frame) \cite{QN}
\begin{eqnarray}
H_{{\rm eff}}(t)= \sum_{i=1}^4 H_i(t) -\delta
(a_1^{\dagger}a_1+a_2^{\dagger}a_2)\nonumber\\
-i\kappa(a_1^{\dagger}a_1+a_2^{\dagger}a_2+2
a_2^{\dagger}a_1)-i\kappa_1^\prime a_1^{\dagger}a_1+
\kappa_2^{\prime\dagger}a_2)
\label{Heff}
\end{eqnarray}
where $a_i$ is the annihilation operator for a photon in cavity $i=1,2$,
$\delta$ is the Raman detuning, $\kappa$ is the decay rate of each
cavity, and $\kappa_{1,2}^\prime$ are the photon loss rate (in mirrors
and transmission channel). This corresponds to the usual one photon
damping due to a zero temperature reservoir. A quantum jump amounts to
the application of the operators $a_{1,2}$.
The Hamiltonian $H_i$ for $i=1,2,a,b$ describes the interaction of the
atoms with the respective cavity modes and the laser, 
where the upper level of the $\Lambda$ has been eliminated
adiabatically already:
\begin{eqnarray}
H_i(t)&=&\frac{g^2}{\Delta+i\Gamma/2}a_i^{\dagger}a_i|r\rangle_i
\mbox{}_i\langle r|
+\delta\omega_i(t)
|e\rangle_i \mbox{}_i\langle e| \nonumber\\
&&-i\left[g_i(t)|e\rangle_i \mbox{}_i\langle r|-h.c.\right],
\end{eqnarray}
where $g$ is the coupling constant between atom and cavity mode,
$\Delta$ is the laser detuning from the upper state in the $\Lambda$
scheme, $\delta\omega(t)=\frac{1}{4}\Omega(t)^2/(\Delta+i\Gamma/2)$
describes both the AC-Stark shift and an effective decay of
$|r\rangle$, $g_i(t)= \frac{1}{2}g\Omega_i(t)/(\Delta+i\Gamma/2)$ is
the effective two-photon Rabi frequency (which is complex), in terms
of the one-photon Rabi frequency $\Omega_i(t)$. In \cite{QN} it is
described how one constructs the proper laser pulse. The quantum jump
corresponding to spontaneous emission amounts to applying the
projection operator onto the state $|r\rangle$ [See Fig.~1(b)].

\begin{figure}[htbp]
  \begin{center}
   \leavevmode
      \epsfxsize=8cm  \epsfbox{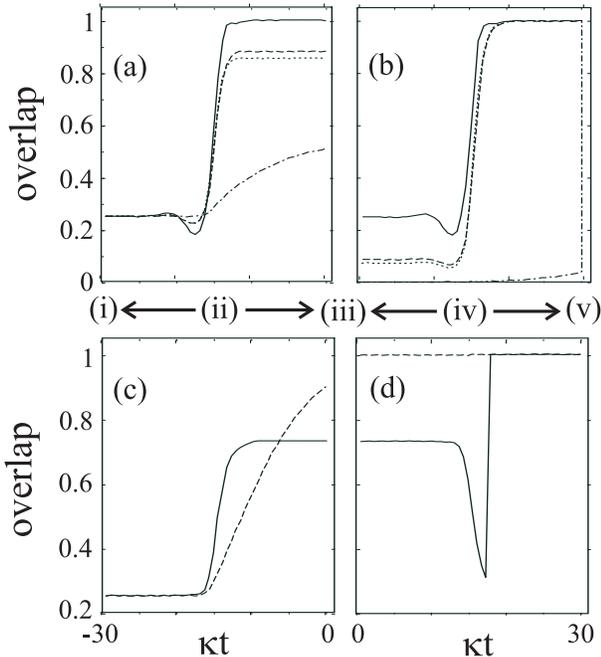}  
\caption{ Results of Monte--Carlo wave function simulations for the
cases of no quantum jump (a,b) and a quantum jump due to photon
absorption (c,d).  The parameters are: $c_0/
\protect\sqrt{2}=-0.29+0.25i$, $c_1/\protect\sqrt{2}=0.36+0.473i$,
$g=5\kappa$, and $\Delta=10\kappa$: (a) Overlap of the state of the
system with the ideal state after step (ii) as a function of time for
$\Gamma=0$ and $\kappa'/\kappa=0,1,10$ (solid, dashed, and dot--dashed
line, respectively), and $\Gamma=\kappa'=\kappa$ and a 10\% error in
the Rabi frequencies $\Omega_{1,2}(t)$ (dotted).  (b) As in (a) but
the overlap with the ideal state after step (iv).  (c,d) Overlap with
the state $c_0|g\rangle_b + c_1 |e\rangle_b$ of the back--up atom in
the case that a jump occurs during step (iv). Plotted are the cases
where $\Gamma=0$, and $\kappa'/\kappa=1,10$ (solid and dashed lines,
respectively).  }
\end{center}
\end{figure}

Figs.\ 2a and b illustrate the time evolution of the full system in
the case that no jump occurs, and where in the final measurements no
error was found (we checked that if an error is found, atom $b$ is in
the correct back-up state).  Fig.\ 2a shows the first half of the
evolution, where we plot the overlap with the ideal state after step
(ii): thus, if there would be no errors, after step (ii) the overlap
would be 100\%.
Fig.~2b shows similarly the
overlap with the ideal final state.  In the absence of absorption and
other errors, both the first and the second gate are indeed found to
transfer 100\% of the population to the desired (intermediate or
final) state.  With increasing absorption, there is
less and less overlap with the correct state after the first gate,
but nevertheless,
the second gate completely recovers from this error (for large
dissipation this happens only in the final step, in the joint
measurement of atoms 2 and $a$), thanks to the 'symmetrization' of
step (iii).  We
also plot a case where there is spontaneous emission and a 10\% error
in the laser pulses, and also there the correct final state is
reached.

Figs.\ 2c and d, show a case where a jump occurred in step (iv),
where now the 
overlap of state of the back-up atom with 
$c_0|e\rangle_b+c_1|g\rangle_b$ is displayed. 
The graph shows that, once the jump occurs, atom 2
will be in that state, and will remain there, also during the
remaining operations, so that the initial qubit is fully restored.

Finally, instead of using the language of quantum trajectories, the
problem can be formulated by including the environment explicitly.
Let us denote by ${\cal T}_{i,j}^{\rm er}$ the unitary transformation  
for the transmission from atom $i$ to $j$. This corresponds to the
state mapping \cite{GG}
\begin{mathletters}
\begin{eqnarray}
|g\rangle_i|R\rangle_j |\xi\rangle_E &\rightarrow&
|g\rangle_i|R\rangle_j |\alpha\rangle_E,\\ |e\rangle_i|R\rangle_j
|\xi\rangle_E &\rightarrow& |r\rangle_i|E\rangle_j |\beta\rangle_E +
|r\rangle_i|R\rangle_j |\gamma_1\rangle_E \nonumber\\ &&+
|e\rangle_i|R\rangle_j |\gamma_2\rangle_E.
\end{eqnarray}
\end{mathletters}
where $|\rangle_E$ denotes a state of the environment including the
cavity modes.  In particular, $|\xi\rangle_E$ is the initial state,
and
\begin{mathletters}
\label{GZ}
\begin{eqnarray}
|\alpha\rangle_E &=&{}_i\!\langle g|\;{}_j\!\langle R| {\cal
T}_{i,j}^{\rm er} |g\rangle_i|R\rangle_j |\xi\rangle_E \equiv T
|\xi\rangle_E,\\ |\beta\rangle_E &=&{}_i\!\langle r|\;{}_j\!\langle E|
{\cal T}_{i,j}^{\rm er} |e\rangle_i|R\rangle_j |\xi\rangle_E \equiv S
|\xi\rangle_E,
\end{eqnarray}
\end{mathletters}
and analogously for the other states of the environment. Ideal
transmission can be accomplished again following steps (i)--(v). The
condition that $\alpha$ and $\beta$ remain the same in the steps (ii)
and (iv) is replaced by $ST|\xi\rangle_E =TS|\xi\rangle_E$. This is
fulfilled, for example, when the Markov approximation applies: this
means that we effectively couple to independent reservoirs in the
first (ii) and second (iv) transmission.

In conclusion, we have presented a scheme which achieves {\em perfect}
transmission in a ``noisy'' quantum network via photon exchange. The
protocol corrects the dominant errors that occur in a physically
realistic situation. The distinguishing feature of our scheme is that
one can repeat the transmission operation as many times as needed to
accomplish ideal transfer. We believe that this is a fundamental
theoretical result towards implementing experimentally quantum
communication networks.

We thank D. DiVincenzo, H.J. Kimble, and H. Mabuchi for discussions.
This work was supported in part by the TMR network
ERB--FMRX--CT96--0087, and by the Austrian Science Foundation.




\begin{references}

\bibitem{QCrypto}
C. H. Bennett, Phys. Today, Vol. 24 (October 1995). 

\bibitem{Schumacher}
B. Schumacher, Phys. Rev. A {\bf 45}, 2614 (1996).

\bibitem{Purification}
C.H. Bennett {\it et al}, Phys. Rev. Lett. {\bf 76}, 722 (1996);
A. Ekert and C. Macchiavello, {\it ibid}, {\bf 77}, 2585 (1996).

\bibitem{densec}See also C.H. Bennett and S.J. Wiesner, Phys. Rev. Lett. {\bf
69}, 2881 (1992); K. Mattle {\em et al.}, {\em ibid.}
{\bf 76}, 4656 (1996).

\bibitem{QN} J.I. Cirac {\em et al.}, quant-ph/9611017.

\bibitem{errorcor}
P.W. Shor, Phys. Rev. A {\bf 52}, 2493 (1995).
A.M. Steane, Phys. Rev. Lett. {\bf 77}, 793 (1996);
E. Knill and R. Laflamme, Phys. Rev. A (in press);
J.I. Cirac, T. Pellizzari, and P. Zoller, Science {\bf 273}, 1207 (1996).

\bibitem{qcerror} To achieve first-order error correction, five
physical qubits are required. R. Laflamme {\it et al.}, Phys. Rev. Lett.
{\bf 77}, 3240 (1996)

\bibitem{Kimble}
Q. Turchette {\it et al.}, Phys. Rev. Lett. {\bf 75}, 4710 (1995).

\bibitem{Iontrap}J.I. Cirac and P. Zoller, Phys. Rev. lett. {\bf 74}, 4091
(1995); C. Monroe {\em et al.}, Phys. Rev. Lett. {\bf 75},
4714 (1995). 


\bibitem{Interf}
We do not allow for spontaneous emission back to the state $|e\rangle_i$
or $|E\rangle_j$. This could be suppressed, for example, by driving
a forbidden transition with the laser, or by tuning
the laser between two excited atomic states to cancel spontaneous
emission by destructive interference, by  
N. L\"utkenhaus {\em et al.}, unpublished.

\bibitem{review}P. Zoller and C.W. Gardiner, in {\em Quantum
Fluctuations}, Les Houches, eds.\ E. Giacobino {\em et al.}, Elsevier,
in press.

\bibitem{cascade}C.W. Gardiner, Phys. Rev. Lett. {\bf 70}, 2269
(1993); H.J. Carmichael, {\em ibid.} {\bf 70}, 2273
(1993);

\bibitem{lpulses}
This means, for example, that we synthesize a laser pulse, duplicate and
time delay it, and use the {\em identical} but {\em imperfect} pulses in
the first and second transmission.


\bibitem{TP1} T. Pellizzari {\em et al.}, Phys. Rev. Lett. {\bf 75},
3788 (1995).

\bibitem{DVNZ}
In our protocol, ideal transmission is achieved eventhough our noisy
channel has a zero probability for no error (i.e. $\gamma_{1,2}\ne 0$).
The capacity of noisy channels with finite probability of no errors has
recently been studied C.H. Bennett, D.P. DiVincenzo and J.A. Smolin,
quant-ph/9701015.

\bibitem{GG} This transformation can be generalized to include
additional atomic states, which will require additional measurements.


\end{references}
\end{document}